\documentclass[vecphys]{svmult}

\usepackage{makeidx}         
\usepackage{graphicx}        
\usepackage{multicol}        
\usepackage{cite}            
\usepackage[bottom]{footmisc}
\usepackage{mathtools,amsmath,amssymb}

\makeindex             

\begin{document}

\title{Magnetization Dynamics}
\author{Andrew D. Kent$^1$ \and Hendrik Ohldag$^{2,3}$ \and Herrmann D{\"u}rr$^{4}$ \and Jonathan Z. Sun$^5$}
\institute{Center for Quantum Phenomena, Department of Physics, New York University, New York, USA (\texttt{andy.kent@nyu.edu}) 
\and SLAC, Stanford University, California, USA   (\texttt{hohldag@slac.stanford.edu})
\and Present address: Advanced Light Source, Lawrence Berkeley National Laboratories and Department of Physics, University of California Santa Cruz, California, USA
\and Department of Physics and Astronomy, Uppsala University, Sweden
(\texttt{hermann.durr@physics.uu.se})
\and IBM T. J. Watson Research Center, Yorktown Heights, New York, USA   (\texttt{jonsun@us.ibm.com})
}

\maketitle

\begin{abstract}

Magnetism primarily describes the physics and materials science of systems presenting a magnetization -- a macroscopic order parameter characterizing electron angular momentum. The order parameter is associated with the electronic exchange interactions, which is fundamentally quantum mechanical. Its dynamic behavior bridges the macroscopic, and the microscopic worlds. On macroscopic length and time-scales, it interacts with electromagnetic fields dictated by the Maxwells equations. On a microscopic scale, it involves the quantum-mechanical electronic states both in spin-space and momentum space, thus giving rise to a wide range of behavior that extend down to femtoseconds. Thanks to the development of modern metrology, there have been many new and noteworthy observations of magnetism-related phenomena across the entire range -- from spin-torque induced antidamping dynamics, to ultrafast laser induced femtosecond electron dynamics that involve spin current and angular momentum conservation.  In this review we introduce some observations on magnetodynamics, and the scientific subjects these new results give rise to. 

\end{abstract}
\setcounter{footnote}{0}
\section{Introduction}

Magnetodynamics spans the entire electromagnetic spectrum. At low-energies and low-frequency, the macroscopic dipolar interaction between a ferromagnetic moment and the long range electromagnetic force is the foundation for all modern electromagnetic power conversion technology. These typically involve time-scales above a millisecond where electromagnetic induction and its interaction with mechanical forces are the primary concern. At shorter time-scales between a millisecond and a nanosecond, one enters the so-called micromagnetics regime, which characterizes the collective motion of magnetic moments under an externally applied magnetic field and its own magnetic anisotropy field. These are still relatively low-energy processes with a characteristic time-scale of the order $\tau_0 \sim 1/\gamma H_\textrm{eff}$ where $H_\textrm{eff}$ is the combined effective magnetic field, typically of the order of the saturation magnetization of a ferromagnet, or a few Bohr magnetons per unit cell, and $\gamma$ is the gyromagnetic constant. Typically such an $H_\textrm{eff} \sim $ several Teslas. This gives a characteristic frequency of approximately 28 GHz/Tesla, where the moment-to-moment interaction is usually dominated by the long-range dipolar magnetic field. On lengthscales below a few hundred nanometers, and the time-scale around nanosecond or less, the exchange interaction between neighboring magnetic moment becomes more significant. But for macroscopic dynamics, the moment still behaves collectively and essentially in a classical manner. In this regime, from long-range dipolar interactions to long wavelength exchange energy\footnote{``long wavelength exchange energy'' in reference to low-energy spin wave modes (magnons) far from zone-boundaries.}, the magnetodynamics can be represented by the phenomenological Landau-Lifshitz-Gilbert, or LLG, Equation\cite{LLG1,Kittel1}:
\begin{equation}
\dfrac{d\bf{M}}{dt} = -\gamma \left ( \bf{M}\times\bf{H}_\textrm{eff}  - \dfrac{\alpha}{\left | \bf{M} \right |}\bf{M}\times\frac{d\bf{M}}{dt}  \right )
\label{eq:LLG} \end{equation}
where  $\gamma = \left | g \right | \mu_B/\hbar$ is the magnitude of the gyromagnetic ratio. $g\approx -2$.  $\mathbf{M}$ is the magnetization,  $\mathbf{H}_\textrm{eff}$ is the local effective magnetic field, including applied and dipolar fields, and includes a derivative of $\mathbf{M}$ if dimensions approach that of the exchange-length, as discussed below in Eq.\ref{JZSE1}. Here $\alpha$ is a dimensionless phenomenological quantity describing damping -- analogous to mechanical friction, or the inverse of a Q-factor used in describing a classical resonator.

In the limit of magnet size small compare to the exchange length\cite{2011044} $\lambda_\textrm{ex} \sim \sqrt{ {{{A}/{K_u}}}}$ where $A$ is the exchange-energy and $K_u$ the anisotropy energy density, the system can be approximately viewed as having a single macrospin moment described by two degrees of freedoms -- i.e. its rotation angles.

Micromagnetics based on the LLG equation remains the most practical methodology for  understanding dynamics in applications today-- applications such as for thin films in magnetic storage (hard disk and tapes) as well as in solid-state integrated magnetic random access memories (MRAMs). 

For time-scales on the order of pico-seconds or shorter, with energies above tens of milivolts per unit cell, the atomic nature of the electronic spin becomes a significant factor. This is the regime where the electronic band-structure and its spin-space degrees of freedom entangle. These interactions lie at the heart of ferromagnetism. The two most important interactions are (1) Heisenberg exchange, originating from the Fermionic nature of the electrons that dictates the occupancy of its atomic orbital states and related band-structures, and (2) a usually smaller spin-orbit interaction, which also plays a role, particularly in materials with heavy elements. This is the regime where electronic transport process gets coupled to that of electron spin degrees of freedom. It gives rise to such properties as the magnetoresistance and anomalous Hall effect in ferromagnetic conductors, where the magnetic configuration affects the electrical transport.  

Over the last two decades or so we began to understand that there exist also an inverse process -- namely, that a transport electronic current carried by itinerant electrons with spins can affect the magnetic state of the conductor as well. For magnetodynamics, this interaction with transport current adds some new terms to the LLG equation which describe the so-called spin-torque effects. Such effects couple magnetodynamics and electronic transport current beyond the dipolar-inductive interaction. These understandings highlight the importance of angular momentum conservation in the detailed balance of spin-polarized electronic transport. 

The recent ability to control magnetization on ultrafast time scales (pico or femtoseconds) and short length scales (nanometers), have attracted interest as such processes may play an important role for information storage and processing technologies in the future. For these, classical approaches to describe such systems are often insufficient, and it is necessary to describe and characterize the magnetic properties of these physical systems on an atomic scale. 

For time-scales much shorter than a pico-second, most of the time one would need a quantum mechanical picture of electronic excitation. The combined electron-spin system can often be modeled by an atomistic approach\cite{2016216}, which uses an LLG equation as a semi-classical approximation on an atomic scale, with spins of fixed length that are exchange coupled via a Heisenberg term. Such models can numerically describe fast dynamics such as ultra-fast laser induced demagnetization.

\begin{figure}
	\begin{center}
		\includegraphics[width=75mm]{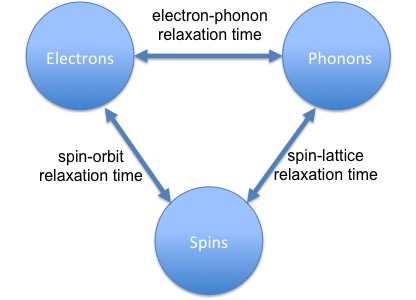}
		\caption{Energy reservoirs in a ferromagnet. Exchange between these is limited by fundamental time constants. Electron phonon relaxation times are around 1-10 ps, spin lattice relaxation times are of the order of 100 ps and spin-orbit relaxation times are of the order of 100 fs.}\label{fig:reservoirs}	
	\end{center}
\end{figure}

An intuitive way to understand excitations involving magnetism is to view the system as composed of a few ``energy reservoirs'' with characteristic coupling strengths (i.e. time-scales), as illustrated in Fig.\ref{fig:reservoirs}. While this as often is an over-simplification it highlights the important factors at play, and the approximate time and energy for the various leading-order factors. In principle it is possible to excite any of these reservoirs through pulsed or periodic optic, magnetic or electric fields. For example, the spin-orbit coupling energy is typically a few tens of meV. If one applies a very strong and short field pulse that is able to overcome this energy barrier on an atomic scale, it will only take about 100~fs for spin and orbital momentum to realign, the spin-orbit relaxation time. Effect on these time scales become relevant when the magnetization is excited using, for example, ultrafast laser pulses that are able to perturb the electron system. However, in cases where external magnetic fields or current pulses are applied the typical electron-phonon relaxation time of 1-10~ps and spin-lattice relaxation time of typically $\approx$~100~ps plays a more important role since such excitation are limited to frequencies of a few tens of GHz compared to optical laser excitations.

In what follows we discuss several recent topics in magnetodynamics that become important thanks both to the advances of fundamental understanding, and to the significant development of modern metrology in X-ray and laser pump-probe measurement technologies. We start with a discussion on the phenomena of spin angular momentum transfer, and the basic nonlinear dynamics involved, which affects ferromagnetic systems in the form of a new term in the LLG Equation. The presence of spin-torque brings us a new, spin-current induced switching mechanism for nano-structured magnets. It also brings a rich variety of non-linear magnetodynamics that have been most directly probed by the recently developed X-ray spin-resolved microscopy. These new discoveries highlight the importance of spin angular momentum current in any such dynamics. We will also briefly review the dynamics occuring on shorter time-scales, often investigated by using femto-second laser excitations.  A more extensive review on the femto-second level magnetic excitation can be found in a separate chapter by T. Rasing {\it et al.} in this volume.

\section{Spin-transfer torque, magnetic switching and oscillations}
\label{JZS}

Spin-transfer torque (STT) is a term that describes the dephasing effect of a spin-polarized current upon entering a magnetically ordered material at a well-defined interface. The dephasing of the spin component transverse to the magnetization under the usually strong s-d exchange-like interaction results in a net torque being exerted on the ferromagnetic moment. The effect was first proposed in spin-valve-like structures \cite{98114,98108} as a manifestation for a type of energy-nonconservative torque discussed in a ferromagnetic tunnel junction \cite{2005073}. Such an energy-nonconserving torque acts along the same vector axis as the damping torque. With the right orientation of the spin-polarization and the spin-current direction, the STT can overcome the damping torque in a nanomagnet, and result in a magnetic reversal if the nanomagnet is situated in an uniaxial anisotropy potential\cite{98114,98128,99078,2000021}. The same torque can also, under more general energy landscapes for the magnetic moment, promote magnetic instability that result in persistent magnetic precession or emission of spin-waves\cite{98108,98130,343}.

\subsection{Spin-transfer torque in phenomenological form}
\label{JZS-1}

Spin-current induced magnetodynamics can in most situations be treated approximately as a torque exerted on the magnetic moment. This is due to a separation of time-scales -- the time-scale involved in conduction electron spin's dephasing is of the order of s-d exchange energy (of the order 0.1 $\sim$ 1 eV, $\sim 1$ fs), and is generally much shorter (of higher energy scale by several orders of magnitudes) than the time-scale involved in typical ferromagnetic material's magnetodynamics in a few Tesla (of the order of $0.1$ ns) of combined anisotropy, dipolar and external magnetic fields.
 
Phenomenologically and in macrospin limit, the STT can be described as an additional term to the Landau-Lifshitz-Gilbert equation, in the form of
\begin{equation}
\frac{1}{\gamma} \left ( \frac{d \vec{n}_m}{d t}\right ) = \vec{H}_\mathrm{eff}\times \vec{n}_m + \left (\frac{\alpha}{\gamma}\right ) \vec{n}_m \times \frac{d\vec{n}_m}{d t} - \left ( \frac{I_s}{\gamma m} \right )\vec{n}_m\times \left ( \vec{n}_m\times \vec{n}_s\right )
\label{JZSE1}\end{equation}
with $\vec{n}_m = \vec{m}/m$ as the time-dependent magnetic moment's unit vector direction, $\vec{H}_\mathrm{eff}$ the total effective magnetic field on the moment (including anisotropy, dipolar demagnetization, and applied field), $\alpha$ the LLG damping coefficient, and $I_s$ the spin-current, here written with a practical unit of magnetic moment/sec (emu/sec), and $\vec{n}_s$ the spin-polarization unit vector direction. $\gamma = \left | g \right | \mu_B/\hbar$ is the magnitude of the gyromagnetic ratio constant. $\mu_B$ here is the Bohr magneton, and $g\simeq 2 $ for simple ferromagnetic metals is the Land\'e g-factor. For a charge current $I_{cg}$ with spin-polarization factor of $0<\eta<1$, for example, $I_s = \left ( \mu_B/e\right )I_{cg} \eta$.

When the magnetic body in volume $\Omega$ is larger than the relevant magnetic length and the system is not in its macrospin limit, the quantities in Eq.\ref{JZSE1} take on a position dependence in addition to the time dependence. Then Eq.~\ref{JZSE1} becomes a {\it constitutive} equation that, with appropriate boundary conditions, determine the micromagnetic dynamics of the system, with the usual non-local  dipolar field interactions described by $\vec{\mathcal{H}}_d\left (\vec{r},\vec{r'} \right )$. The location dependence of the magnetic moment also introduces an exchange-energy related term into Eq. \ref{JZSE1}\cite{2008117}. These result in a modification of the field term as
\begin{equation}
\vec{H}_\mathrm{eff} \rightarrow \vec{H}_\mathrm{eff}\left ( \vec{r} \right ) + \int_{\Omega}\vec{\mathcal{H}}_d\left (\vec{r},\vec{r'} \right )d^3r' + \left ( \dfrac{A}{M_s} \right ) \bigtriangledown^2 \vec{n}_m\left ( \vec{r}\right )
\label{JZSE2}\end{equation}
which describes the long-wavelength effect of ferromagnetic exchange, with $A$ being the magnetic exchange energy, and a replacement in Eq.\ref{JZSE1} of a macrospin direction vector with a unit vector for the direction of the local saturation magnetization, that is $\vec{m}\rightarrow \vec{M_s} = M_s\vec{n_m}\left ( \vec{r} \right ) $.

The source of interface spin-current $I_s$ may include spin-filtering of a spin-polarized current in metals, such as the Valet-Fert type of current-perpendicular-to-plane (CPP) spin-valve structures, either diffusive or ballistic\cite{2002116,2002009}. It may originate from spin-filtering through a barrier in magnetic tunnel junctions (MTJs)\cite{2004033,2005129}. The spin-current could also be generated by transport processes involving strong spin-orbit interaction\cite{2007003,2007002,2012117,2012040,2016111,2010030}. Large spin-torques have also been reported at interfaces between topological insulators and ferromagnets, see Refs. \cite{2014061,2011198,2012142}, for example, for a review on the subject. The spin-current can also be induced by heat-flow involving a magnon thermal-gradient\cite{2010088,2016112,2016113}. For these non-filtering mechanisms, the electron to spin current conversion is in principle not limited to one $\hbar/2$ per electron charge current, whereas for the spin-filtering processes, one electron can only transfer one $\hbar/2$  of spin-angular momentum, that is one Bohr Magneton $\mu_B$ of magnetic moment, which sets a limit to the charge-spin conversion efficiency.

Two factors make STT-dynamics rich and often complex. They are: (1) the effect of finite temperature, which makes thermal fluctuations, and thus related stochastic distribution issues important for STT-driven dynamics, and (2) in most realistic samples various internal degrees of magnetic freedom participate in the STT-dynamics, making the whole process more complex than a macrospin. Both are significant scientific challenges to one's ability to quantitatively understand and control the STT-related behavior. Both are critically important for technology development. 

\begin{figure}
 	\begin{center}
 		\includegraphics[width=4.5in]{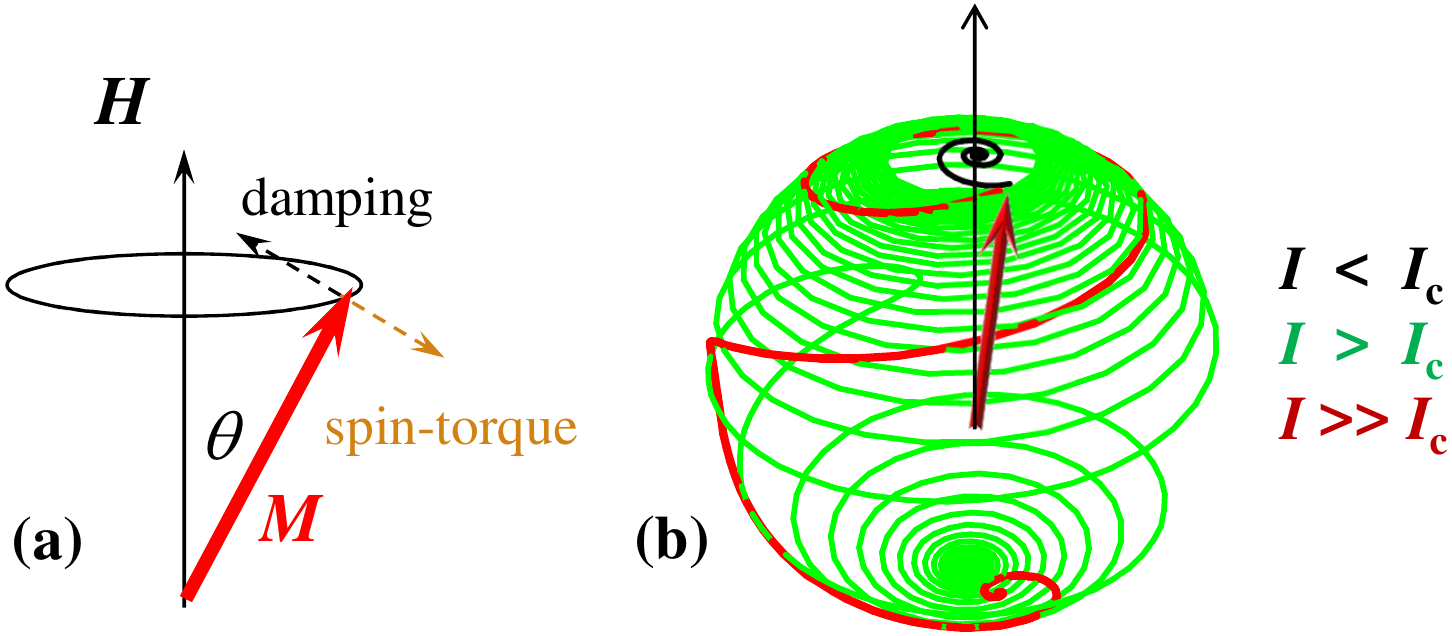}	
 		\caption{An illustration of the dynamics involved in spin-torque induced magnetic switching. (a) The vector torque relationship. The de-phasing induced spin-torque has the same axial direction as the damping torque, in the direction of $\mathbf{n}_m\times \left ( \mathbf{n}_m\times \mathbf{n}_s\right )$ as described in Eq.~\ref{JZSE1}. It can be parallel or antiparallel to the damping torque depending on the sign of the spin-current. (b) The trajectory of a macrospin on its unit-sphere under the combined influence of damping and spin-torque. When spin current is less than the threshold $I_c$ necessary for net zero-damping, the moment damps towards its north-pole energy minimum. When spin current exceeds threshold $I_c$, an anti-damping precession ensues in the northern hemisphere, with cone angle $\theta$ increasing over time. The rate of cone-angle opening is faster for larger spin-current.}
\label{Fig:STT_Sphere}
\end{center}
 \end{figure}

The primary effect of spin-torque is to change the apparent magnetic damping $\alpha$ of the receiving magnetic nanostructure. When a sufficient amount of spin-torque is applied in the right direction, the net damping of the nanomagnet can become negative for a certain region on the macrospin coordinate unit-sphere, which would result in amplifying magnetic precession, as illustrated in Fig.~\ref{Fig:STT_Sphere}. For a nanomagnet situated in a uniaxial potential, this could result in its moment reversing from the easy-axis direction anti-aligned to the spin-polarization to the direction aligned\cite{98114,98128,2000021}. In other geometries, it is possible to send the moment into persistent oscillation mode, which becomes an STT-driven oscillator\cite{2003085}.

\subsection{Spin-torque driven anti-damping magnetic switching }
\label{JZS-2}

Besides its novel anti-damping magnetodynamics, STT-driven magnetic switching is technologically interesting primarily because it can effectively manipulate and switch a nanomagnet situated in an anisotropy potential. A nanomagnet in an uniaxial anisotropy potential with at least two stable moment-direction states is the building block of a magnetic memory bit. The STT-mechanism provides a {\it local current-based} addressing mechanism, circumventing the long-range nature of magnetic field-based write-mechanisms such as a magnetic write-head, thus enabling high-density, all solid-state magnetic memory (STT-MRAM). 

For a macrospin, from Eq.\ref{JZSE1}, the instability threshold for a collinearly aligned anisotropy axis, applied field direction and spin-current polarization direction, the threshold switching spin-current is simply 
\begin{equation}
I_{c0} =\left ( \frac{2 e}{\hbar} \right ) \left (\frac{\alpha}{\eta} \right ) m H_\mathrm{eff}
\label{JZSE3}\end{equation}
where $\eta$ is the effective spin-polarization of the charge current, whose exact functional form depends on the structure of the nano-magnet's spin-current transport environment\cite{98114,2002123,2005129,WileySlonczewski08,2016084}. $H_\mathrm{eff} = H_k + H_a$ in the case of a simple uniaxial anisotropy with its anisotropy field as $H_k$ and the applied field $H_a$ in the same direction. 

At low temperature $T\rightarrow 0$ K limit, a spin-torque above $I_{c0}$ is required to initiate the anti-damping switching. In macrospin limit, the time it takes to switch a nanomagnet in the macrospin-limit is inversely proportional to $I - I_{c0}$ in the form of 
\begin{equation}
\tau \approx \dfrac{\tau_0}{I/I_{c0}-1} \ln \left ( \dfrac{\pi}{2 \theta_0 }\right )
\label{JZSE3B}\end{equation}
with $\theta_0$ being the intial angle the moment makes with its easy-axis upon the switching-on of the spin-torque, and $\tau_0 = \dfrac{m/\mu_B}{\eta I_{c0}/e} = 1/\gamma \left (H_k+H_a \right ) \alpha$\cite{2000021,2014031,2016084}. 

At finite temperature there is a thermal distribution of initial angles $\theta_0$ about the easy-axis. The precessional reversal process would also be thermally scattered. Of these two processes, in the short time, high-STT drive limit  (strictly speaking in the asymptotic limit of $I/I_{c0}\gg10$\cite{2014031}) where $\tau \ll 1/\alpha \gamma H_k$, the initial condition's thermal distribution usually dominates the consideration\cite{2014031,2013041}. Taking this distribution into account, one arrives at the probability for switching at time $t$ for $I\gg I_{c0}$ as $P\left ( t \right )$, which has the high-barrier $E_b\gg k_B T$ and long-time $t\gg \tau_0/\left (I/I_{c0}-1 \right )$ asymptote as\cite{IBMmemo2006,2007056,2014031,2014011}
\begin{equation}
P\left ( t \right ) \approx \exp \left[ -\frac{\pi ^{2}E_b }{4 k_B T}e^{-\frac{2t}{\tau _{0}}\left ( I/I_{c0}-1 \right )}\right] +O\left[ \exp
\left( -\dfrac{\pi ^{2}E_b }{4 k_B T}\right) \right] 
\label{JZSE4}\end{equation}
for the limit of $ I\gg I_{c0}$ and $ P\left ( t \right ) \rightarrow 1$.

In the sub-threshold $I<I_{c0}$ region, while no switching would be expected at zero temperature, with finite temperature, thermally-assisted reversal has a finite probability. This switching probability, however small it might be, will also be magnified by the application of spin-torque, the resulting sub-threshold switching probability in macrospin-limit is given by 
\begin{equation}
P\left ( t \right ) \approx 1-\exp\left\{ -\gamma_0 t \exp\left [ -\dfrac{E_b}{k_B T} \left (1-\dfrac{H}{H_k} \right )^{\nu_1} \left (1-\dfrac{I}{I_{c0}}\right )^{\nu_2} \right ] \right\}
\label{JZSE5}\end{equation}
for $0 \lesssim P\left ( t \right ) \ll 1$ with $I \ll I_{c0}$ and $H \ll H_k$.  $\nu_1 \lesssim 2$ if $\vec{H}$ is nearly collinear with the anisotropy axis, otherwise $\nu_1<2$\cite{2011002B}, and $1\leq \nu_2 \leq 2$ depending on the details of the anisotropy potential shape\cite{2013115,2013109}, with $\nu_2\rightarrow 2$ for a perfectly collinear, uniaxial-anisotropy-only configuration. The exact values of the exponents $\nu_{1,2}$ in a real-life nanomagnet structure (such as a magnetic tunnel junction) depend further on details of the system's micromagnetics behavior where the structure is usually larger than a macrospin\cite{2015063,2016082,2011136A,2011176,2014035}, and the exponents in such experimental systems can generally vary around the range described above with some uncertainty from device to device, for different materials combinations, for different measurement time-scales and drive amplitudes. 

Eqs.(\ref{JZSE4} - \ref{JZSE5}) form a pair of asymptotic descriptions of the probabilistic switching behavior of a macrospin under collinear spin-torque. Note that neither covers accurately the ``most-likely switching'' region of $I \sim I_{c0}$ or when $P \sim 1/2$. For this cross-over region's accurate mathematical macrospin-solution, a full Fokker-Planck equation-based numerical treatment is often necessary\cite{2004147,2007056,2014011}.

Experimentally, the STT-induced magnetic switch was observed earlier in highly spin-polarized manganite tunnel junctions\cite{98128}, in nanostructured spin-valves\cite{99078}, and in magnetic tunnel junctions (MTJs) \cite{2004033}. The advent of MTJs, especially later the MgO-barrier based MTJs\cite{2005017,2005016,2005130,2006031} with perpendicular magnetic anisotropy\cite{2010058,2011014,2008097} significantly accelerated technological development in memory applications, due to their impedance match into the existing CMOS circuits, their large magentoresistance (MR) for read-out, and the PMA-based magnetic storage bit's reduction to switching threshold current, and its PMA energy density's scalability to very small sizes (with 11nm diameter demonstrated on CMOS integrated wafers\cite{2016085}, and 8nm on individual junctions\cite{2018051} at the time of this writing). 

Early experimental observations of STT-driven switches were made over a time-scale typically longer than a $\mu s$. These at finite temperature only probe the sub-threshold, thermal-activation mediated switching process. The drive-speed dependence of such switching, either with STT- or with field-drive (easy-axis hysteresis measuring $H_c$'s rate dependence) can be used to estimate the thermal activation barrier height\cite{2005098,2002113,2005133,2011136A}. 

Nano-second level super-threshold switching was first observed in metallic spin-valves, demonstrating the characteristic switching-speed and drive-amplitude trade-off similar to Eq.\ref{JZSE3B}\cite{2004183,2004182,2005004,Bedau2010a,Bedau2010b}. A similar fast-switching characteristics was also observed in MTJ-based switching events on back-end CMOS integrated devices\cite{2005133}. Nano-second speed STT-induced switching using non-local spin-current was also seen in spin-valve-like filtered structures\cite{2011001B} and in spin-Hall induced spin-currents\cite{2015112,2016119}.

Fast, reliable switching is essential for memory applications of STT-based devices. To this end a significant amount of effort is directed to experimentally assessing the switching statistics as it depends on the drive voltage's amplitude and duration in a PMA MTJ.  For macrospins in a uniaxial anisotropy potential with barrier height $E_b$, a relatively simple relationship is obtained between the driving spin-current amplitude, drive duration, and the amount of switching error, in the asymptotic form of 
\begin{equation}
I_w \approx \dfrac{E_b}{\kappa} + \dfrac{Q_0}{\tau_w}
\label{JZSE7}\end{equation}
for a write pulse width in the range of $\tau_w \sim \tau_0$ or smaller.  Here, $I_w$ is the write-current amplitude. $\kappa = E_b/I_{c0} \approx \left (\hbar/4 e\right ) \left ( \eta / \alpha \right )$ is the so-called STT ``switching efficiency'' here expressed in macrospin limit\cite{2011136A,2011176}. From Eq.\ref{JZSE4} and Ref.\cite{2011136A}:
\begin{equation}
Q_0\approx \left ( \dfrac{e}{2 \eta } \right ) \left ( \dfrac{m }{\mu_B } \right ) \ln \left ( \dfrac{\pi^2 E_b}{4 k_B T \epsilon_r } \right )
\label{JZSE8}\end{equation}
with $\epsilon_r = 1 - P\left ( \tau_w \right )$ representing the write-error probability.  While the derivation of Eq.\ref{JZSE7} relies on macrospin model, all parameters involved ($E_b$, $\kappa$, $Q_0$, $\tau_w$ and $I_{c0}$) can be experimentally determined. The resulting $I_w$ is to the leading order a viable estimate even beyond the macrospin limit. The departure from macrospin is mainly captured by sub-volume nucleation of thermally activated reversal states\cite{2011136A,2012152,2018008}, which makes the value of $\kappa$ size-dependent and below its macrospin value, and by $Q_0$ being drive-amplitude dependent and smaller than macrospin-value\cite{2016192}.

STT-switched PMA MTJs were demonstrated to switch at 10ns pulse width reliably with a switching error below $10^{-11}$ per switching operation on single device level\cite{2011121}. The most recently demonstrated 10ns switch with deep statistics was on an 11nm diameter PMA MTJ at an error rate of $10^{-9}$\cite{2016085}. Even at 11nm diameter, the device's behavior is still not completely in the macrospin limit, but shows steeper decrease of the switching error-rate upon the increase of pulse height, indicative of fractional volume initiation of the switching process\cite{2016085,2011136A}.

In this aspect of the STT-operation, one challenge for a successful technological deployment of STT-based memory is to obtain high reliability STT-switching (with $\epsilon_r \ll 10^{-9}$ at the very least) with as small a value of $I_w$ as possible while retaining write speed and a sufficient thermal activation barrier height $E_b$ to ensure 10-year nonvolatility of stored data-bit. A review of the development status of STT-MRAM and a comparison of STT-MRAM to other solid-state memory technologies can be found in Refs.\cite{KentWorledge2015,2016192}.

\subsection{Orthogonal spin-torque driven magnetic switching}\label{ADK}

In addition to the anti-damping type of magnetic switching, when sufficiently strong, a spin-torque can result in precessional switching.  This type of switching requires a spin-torque approximately $1/\alpha$ stronger than the anti-damping type of switching, since now the torque needs to drive dynamics associated with the magnetic anisotropy. This mechanism was originally proposed as a high-speed precessional switching method \cite{2004213}. It has also been recently demonstrated with a slight modification\cite{2016122,2016124} in spin-Hall effect-based orthogonal spin-torque 3-terminal type of devices\cite{2010109,2016122,2016123,2016124,2012040,2015147,2014141,2014024}, where the spin-Hall effect induced spin-current was proved to be sufficiently strong, possibly with some help from a `magnetic field-like' component as well due to symmetry-breaking spin-orbit interactions at the interfaces. 

\begin{figure}
 	\begin{center}
 		\includegraphics[width=6cm]{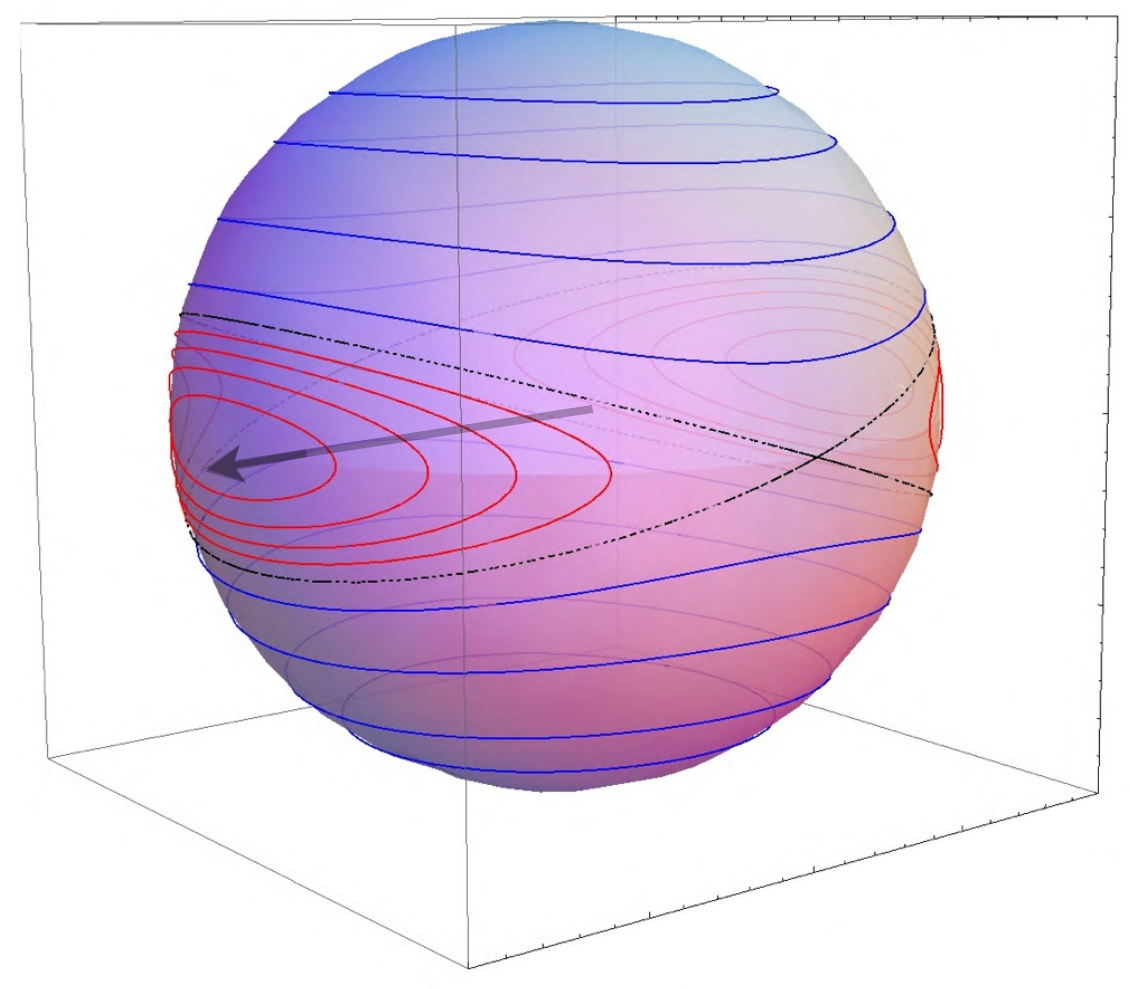}	
 		\caption{Precessional orbits of a biaxial macrospin. The red lines show in-plane orbits about the easy axis ($x$), while the blue lines illustrate out-of-plane magnetic orbits about the the hard axis ($z$). The black line is the separatrix, marking the boundary between in-plane and out-of-plane orbits. Figure adapted from Ref.~\cite{Pinna2014}.}
		\label{Fig:Biaxial}
 	\end{center}
 \end{figure}	

For an in-plane magnetized free layer this requires spins with a polarization component perpendicular to the plane. The free layer then has biaxial magnetic anisotropy with an easy axis in the film plane (typically set by an asymmetric shape of the element, e.g. an ellipse) and a hard axis perpendicular to the plane. There are then two types of precessional orbits, as illustrated in Fig.~\ref{Fig:Biaxial} \cite{Pinna2013,Pinna2014}. One type is around the easy axis and is similar to the orbits in a uniaxial nanomagnet discussed above. The second type are orbits around the hard axis, orbits of the magnetization out-of-the film plane. For a spin-polarization collinear with the easy axis the switching is of the anti-damping type described above, and a spin-torque of the appropriate sign (i.e. spin current polarity) results in a direct switching of the magnetization. However, when there is a sufficient component of spin-polarization out of the film plane out-of-plane orbits can be excited and the switching can be precessional, with a speed set by the demagnetization field ($\tau \sim 1/(\gamma M_s)$). This switching can be very fast ($<50$ ps) but the switching probability will typically be an oscillatory function of the pulse amplitude and duration, meaning that the pulse timing may be critical to obtaining a low write error rate. Switching with current pulses as short has $500$ ps has been demonstrated in MTJ devices that incorporate a perpendicularly magnetized spin-polarizing layer, known as an orthogonal spin-transfer (OST) device \cite{Liu2010,Rowlands2011}, and switching with even shorter current pulses has been reported in spin-valve devices with a perpendicular spin-polarizing layer \cite{Lee2009,Beaujour2009,Papusoi2009,Park2013}.

An exciting recent modeling result demonstrated that an adiabatically decaying current pulse can lead to highly reliable precessional switching \cite{Pinna2016}. The essential reason is that a spin-torque bias can be present, by a slight tilt of the spin-polarization direction away from the exact orthogonal direction of the anisotropy's easy-axis,  as the magnetization relaxes toward one of its easy directions, from the out-plane orbits (blue curves in Fig.~\ref{Fig:Biaxial}) to the in-plane orbits about the easy axis (red curves n Fig.~\ref{Fig:Biaxial}). As of this writing this model remains to be tested in experiment.

\section{Spin-torque oscillators}\label{ADK1-4}

\subsection{Spin transfer induced excitation of spin-waves}
Spin-torque can under many situations cause persistent magnetic oscillations and excite spin-waves. This can either be of the form of coherent precession of a near macrospin \cite{2000021,2003085,2007140} or spatially non-uniform magnetization patterns \cite{98108,98130,98106,2003017,2004007,2005057,2007045} of various wavelengths and amplitudes. In both cases the magnetization dynamics can be highly non-linear.

\begin{figure}[t]
 	\begin{center}
 		\includegraphics[width=5cm]{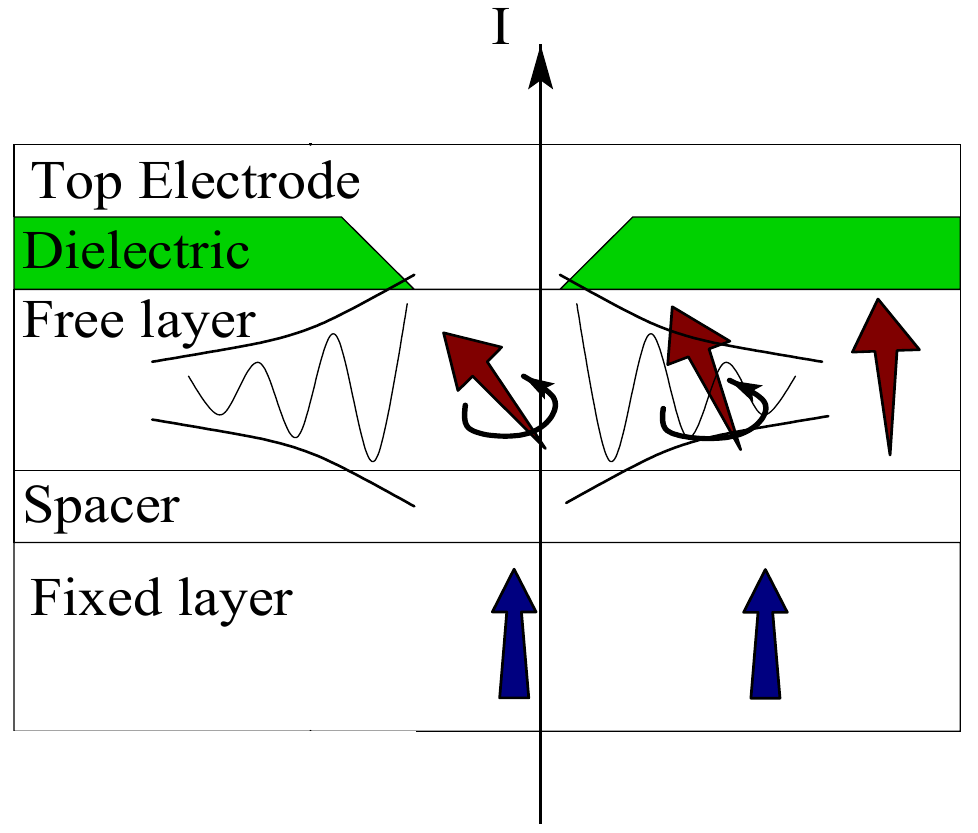}	
 		\caption{Schematic of a lithographically defined nanocontact to a magnetic bilayer. Current flow in the contact produces a STT on the moments in the contact region that excites spin-waves. The spin-waves can propagate away from the contact under certain conditions.}
		\label{Fig:STNO}
 	\end{center}
 \end{figure}	

A point contact to an otherwise extended ferromagnetic thin film stack is a prototypical configuration that can be used to excite spin-waves. The STT is concentrated in the contact region and thus will generate a non-uniform magnetic excitation (i.e. because regions outside the contact do not experience a significant STT). A typical structure consists of a thin film stack in the form $\parallel$ FM $\mid $ NM $\mid$ FM (free) $\parallel$ surface$\parallel$ and the point contact can be either via a metallic tip \cite{98106} or by lithographic patterning \cite{2003017} to create a ``nanocontact,'' an aperture with a diameter $\lesssim 200$ nm in a dielectric that is filled with a non-magnetic metal (Fig.~\ref{Fig:STNO}).

The magnetization excitations can be localized or extended. In the former case spin-excitations are near the contact and decay rapidly outside the nanocontact region. While in the latter case, propagating spin-wave modes can be excited. The nature of the excited modes (extended or localized) depends on many characteristics of the contacts, ferromagnetic layers and applied current and field. The basic issue is whether spins excited by STT in the contact region can couple to the propagating spin-waves in the free layer. 

In a linear response region (limit of low currents, currents just above the threshold for current induced excitations) it was shown theoretically that the wavelength of the spin-waves would be approximately the nanocontact diameter and propagating spin-waves would be excited, with spin-wave amplitudes that decay algebraically with distance from the nanocontact \cite{Slonczewski1999}. However, this situation not generic. At higher current amplitudes spin-waves can be localized in the contact region \cite{Slavin2005,Hoefer2005,Hoefer2008} and the nature of the excitations depends on the applied field (i.e. there can be transitions between localized and extended spin-waves as a function of the applied field \cite{Bonetti2010}).  Both extended and localized excitations have been observed experimentally using optical and X-ray imaging methods, such as Brillouin light scattering \cite{Demidov2010,Madami2011}, Kerr effect \cite{Madami2015} and scanning transmission X-ray microscopy \cite{Backes2015,Bonetti2015}. 

In the case in which the free layer has perpendicular magnetic anisotropy and the field is perpendicular to the layer plane the excitations are localized and spin-waves can condense in the contact region forming a magnetic droplet soliton \cite{Hoefer2010}, a state that was predicted theoretically in the 1970s for a ferromagnetic layer with axial anisotropy and no-dissipation or damping \cite{Ivanov1977,Kosevich1990}. This has recently been observed experimentally \cite{Mohseni2013,Macia2014,Chung2016}, including directly using scanning X-ray transmission microscopy (STXM) \cite{Backes2015}. In the latter experiment the profile of the excitation in the contact region was directly measured and the spin-wave amplitude was seen to decay rapidly outside the contact, i.e. the spin-excitation was shown to be localized in the contact region. Modeling has shown that STT in nanocontacts to thin films with perpendicular magnetic anisotropy can also stabilize topological magnetic structures, including dynamical magnetic skyrmions \cite{Zhou2015}. Solitons were also directly observed in in-plane magnetized free layers in which the Oersted field from the current leads to a more complex potential for spin-waves \cite{Bonetti2015}, for example, the potential favors propagation of spin-waves in particular directions away from the contact. In this case, STXM was used to create movies of the magnetic excitations, using a heterodyne method of locking the frequency of the spin-excitations to an external microwave source. Further, this directional propagation of spin-waves has been used to phase lock multiple nanocontact oscillators \cite{Houshang2016}. 

%

The discovery of very large spin-orbit torques in heavy non-magnetic transition metals (e.g. Pt, Ta and W) \cite{2010109,2012040} has lead to new types of spin torque oscillators. The geometry consists of a heavy metal film with an interface to a magnetic layer; a prototype thin film structure is  
$\parallel$ substrate $\mid$ non-magnetic TM $\mid$ FM (free) $\parallel$.
In these oscillators the charge current flows in the plane of the non-magnetic transition metal layer but spin-current flows perpendicular to the plane creating spin-wave excitations in the free layer (or magnetic switching of the free layer as discussed in Sec.~\ref{JZS-1}). Oscillators based on few micron diameter Permalloy ferromagnetic disks \cite{Demidov2011a,Demidov2012b} and sub-micron in-plane magnetized CoFeB elements \cite{Liu2012C} have been realized. In addition, various contact and ferromagnet layer geometries have been demonstrated \cite{Demidov2014A,Zheng2014,Ranjbar2015}. In addition, the phase locking of multiple spin Hall nanooscllators was recently demonstrated \cite{Awad2015}.

The fact that charge current only flows in the plane of the non-magnetic transition metal and need not flow perpendicular to the plane (as in nanocontact and nanopillar based spin-transfer devices) enables exciting magnetization dynamics in magnetic {\em insulators} using spin-orbit torques. This is interesting for a number of reasons, one of which is that spins waves can in principal travel over much larger distances because magnetic insulators can have orders of magnitude lower damping than conducting magnets.  Another is that spin-orbit torques enable control and modification of the spin-wave spectrum of magnetic insulators. A variety of experimental studies of yttrium iron garnet (Y$_3$Fe$_5$O$_{12}$, YIG) films, a ferromagnetic insulator with very low damping ($\alpha <10^{-4}$), have been reported recently showing spin-orbit torque induced persistent magnetizations oscillations and the excitation of propagating spin-waves \cite{Kajiwara2010,Hamadeh2014,Collet2016}. Spin-excitations in magnetic insulators using spin-transfer torques is a new and rapidly developing area at this time.

\subsection{Spin transfer vortex oscillators}

\begin{figure}[b]
 	\begin{center}
 		\includegraphics[width=5cm]{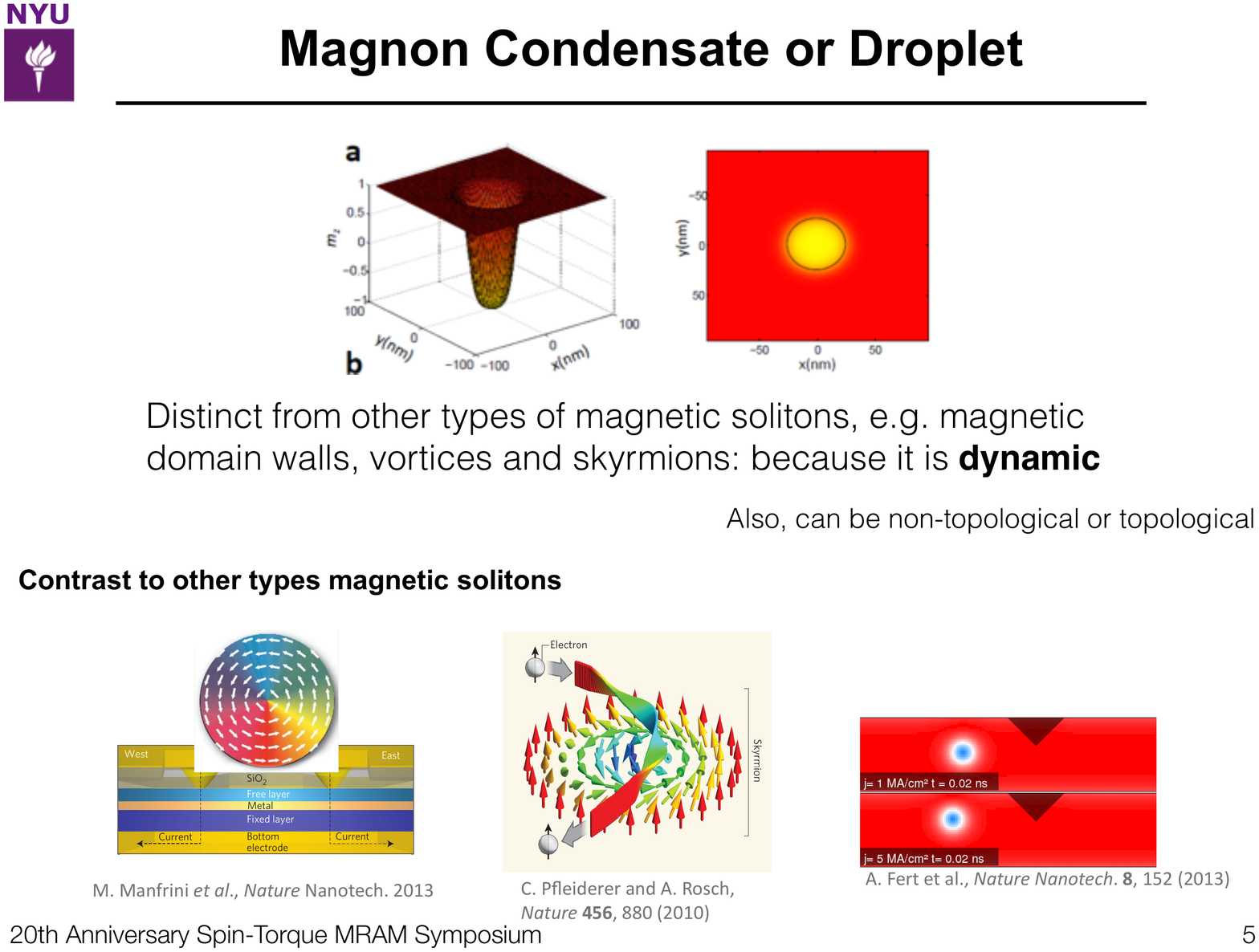}	
 		\caption{Schematic of a nanocontact with the free layer in a magnetic vortex configuration. A pulse current can nucleate a vortex, the planar spin-configuration shown above the contact and currents used to excite gyrotropic vortex. This motion can be studied by measuring the contact resistance as a function of time or spectral measurements. Figure adapted from Ref.~\cite{Manfrini2014}.}\label{Fig:Vortex}
 	\end{center}
 \end{figure}

Spin transfer can also used to nucleate, excite and study magnetic vortices and can be applied to the study of other types of topological magnetic structures (e.g. Skrymions). Vortices can form in thin ferromagnetic layers and disks composed of soft magnetic materials, such as permalloy. In a nanodisk they minimize magnetostatic energy at the cost of exchange and anisotropy energy, forming a circular spin structure shown in Fig.~\ref{Fig:Vortex}. They can thus be favored by the geometry of the magnetic layer, typically thicker disks will have a tendency to have vortex ground state magnetic configurations (see, for example, \cite{Cowburn1999}). In a thin extended layer, such as in a spin-transfer nanocontact (Fig.~\ref{Fig:Vortex}), vortices can be nucleated by applying a current pulse, as the Oersted field from the pulse favors a circular spin-configuration~\cite{Manfrini2014}. The center of the vortex has a singular region, a core, in which the moments point out of the film plane~\cite{Shinjo2000}. There are thus two physical quantities that characterize a magnetic vortex, the sense of spin circulation in the plane (a winding number) and the direction of core magnetization (a polarity). Both these quantities play an important role in the dynamics of vortices---as described by the Thiele equation, an equation derived from the LLG equation that describes a vortex's ``center of mass'' motion as well as that of other types of magnetic objects (e.g. domain walls) \cite{Thi1973}.

Part of the interest in vortex oscillators is that that can have high quality factors \cite{Pribiag2007}. Their oscillation frequency is typically 0.1 to a few GHz, about one order of magnitude less than spin-wave frequencies (i.e. the FMR frequency). However, the oscillator's output power can be large, particularly when the magnetic layer in which the vortex forms is one of the electrodes of a magnetic tunnel junction nanopillar \cite{Dussaux2010}. Spin torques also enabling exciting non-linear dynamics and vortex oscillators are a physics playground for exploring such dynamics, including chaotic magnetization dynamics \cite{Watelot2012}. A variety of experimental studies have examined the interactions between magnetic vortices, including in nanopillars with two magnetic layers that each contain a vortex~\cite{Sluka2015} and in proximal nanopillars~\cite{Locatelli2015}.  Further, as in been demonstrated in other types of spin torque oscillators \cite{Rippard2005}, injection locking of vortex oscillators has been demonstrated~\cite{Burgler2011}. Time resolved imaging studies of magnetic vortex dynamics have been conducted using MOKE \cite{Park2003} as well as X-ray microscopy \cite{Choe2004}.

\section{Synchrotron and femtosecond-laser based time-resolved spin-dynamics}\label{HO}

The previous sections focused on dynamics on the time scale of 10s to 100s of picoseconds. As described in the introduction, electrical transport measurements are the workhorse for experimental characterization in this area. They make use of magnetoresistance effects and suitable electronics operating at tens of GHz can provide information about the macroscopic dynamic magnetic behavior. However, the exact microscopic behavior caused by the interplay of spin torque with the effective field and the exact shape of the element can only be indirectly deduced if the transport results are compared to micromagnetic simulations. For direct observation of many such dynamics involved in spin-torque driven oscillations in nano-structures, the recently developed spin-resolved X-ray microscopy proves to be a useful tool. After describing the different switching schemes that are used in x-ray based and optical experiments we will give a brief review on this type of measurements. A more detailed review on this subject can be found in Ref.\cite{Stoehr2006,UMC13}.

\subsection{Switching Schemes}

\begin{figure}
	\begin{center}
		\includegraphics[width=100mm]{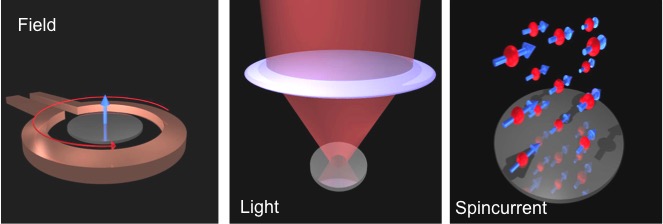}
		\caption{Illustration of different switching schemes, via application of a local magnetic field (left), optical excitation (center) or spin polarized current (right).}\label{fig:switching}
	\end{center}
\end{figure}

In the following we are concerned with the manipulation and ultimately the reversal of magnetization on fast time scales. 
Naively one would expect that the magnetization of a sample reverses upon application of a magnetic field of the opposite direction. However, this is not the case, since the origin of the magnetization is the spin angular momentum of the atoms and to change angular momentum one has to supply angular momentum to the sample. The most common way to do this it to apply a torque with an external field. That means that the external field has to be applied at an angle to the magnetization. This is in general described in the Landau-Lifschitz Gilbert equation shown below in Eq.~\ref{eq:LLG}, which includes the gyromagnetic ratio $\gamma$, the effective field (sum of external field, demagnetization field and anisotropy fields) as well as the damping constant $\alpha$, which is necessary to include dissipative processes in a realistic medium.

Fig.~\ref{fig:switching} shows three different possibilities to affect the magnetization by supplying angular momentum or by applying a torque to the current magnetization in an experiment. The panel on the left hand side shows the simple case of applying a magnetic field. While this approach is rather straight forward to realize and can be described using Eq.~\ref{eq:LLG}, it has practical limitation given by the maximum field that one can achieve and the fact that it is non-trivial to localize magnetic fields to the area of interest. The fact that field induced dynamics on larger length scales are rather complex limits the speed with which the magnetization can be reversed using external fields. Interestingly, one finds that even when using the extremely strong fields of several Tesla generated by a relativistic electron bunch for a picosecond or less, the magnetization far away from the excitation is still precessing for about 100 picosecond \cite{Tudosa2004}, essentially limiting the speed with which information stored in such a system can be processed by only using conventional magnetic fields.

To speed up the reversal and to control the magnetization in a more confined manner other excitation mechanisms are considered. For example, the surprising possibility to manipulate magnetization using short and powerful laser pulses has evolved over the past decade \cite{Hansteen2006,Hansteen2005,Kimel2005,Stanciu07,Lambert2014}. Here a magnetic sample is irradiated with an ultrashort laser pulse without the application of an external field or current. The exact nature of the reversal mechanism and the role of the angular momentum supplied by the photons is still debated. We will give a short overview over this field in section \ref{HD}. A more thorough overview will be given by T. Rasing {\em et al.} in another chapter. 

One of the most efficient ways to detect magnetization dynamics on these time scales is via electrical transport. This approach uses standard radio frequency analysis and lock-in equipment that is commercially available, which is why it is widely used (review). For this detection method one makes use of magnetoresistance effects like giant magnetoresistance (GMR) or anisotropic magnetoresistance (AMR). Any time one investigates a layered magnetic system it is usually possible to use the GMR effect, which manifests itself as a change in magnetization when a current flows between two magnetic layers separated by an non-magnetic layer. This is in particular useful for studies of spin-torque oscillators where the current flows through a small nanocontact from a fixed polarizing layer to a free oscillating layer. In this case one can make use of the already existing electrical contacts and directly determine the magnetization dynamics using a spectrum analyzer and lock-in detection. If there is only one magnetic layer or the devices are simply planar one can use the AMR effect, which manifests itself in a change in resistance if the magnetization is parallel or perpendicular to the current direction. This effect is at least an order of magnitude smaller than the GMR but it is compatible with most experimental geometries. However, none of these approaches provide microscopic or spatially resolved information. Typically, one can obtain indirect information about the microscopic behavior of the sample by comparing the macroscopic transport results to model calculations.

\subsection{Microscopy using Visible Light or X-rays}

\begin{figure}
	\begin{center}
		\includegraphics[width=100mm]{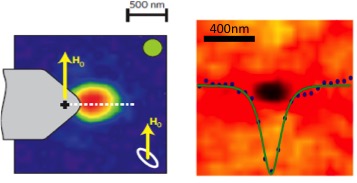}
		\caption{Two different approaches to image spin wave dynamics excited by a local nanocontact. The image on top shows an image acquired using Brillouin Light Scattering from \cite{Demidov2010}, while the image on the bottom shows a localized spin wave acquired using scanning transmission x-ray microscopy \cite{Backes2015}.}\label{fig:imaging}	
	\end{center}
\end{figure}

In order to obtain microscopic information about magnetization dynamics it is possible to either use visible light microscopy or x-ray microscopy. Microscopy using visible light based on the magneto-optical Kerr effect has the advantage that the setup can be fairly simple and and requires few components. Also using standard pump-probe approaches one can achieve very good time resolution of the order of 100 fs. However, the spatial resolution is limited and due to the fact that the Kerr rotation is generally small means that it is not generally possible to use time resolved MOKE microscopy. Nevertheless, using sophisticated lock-in approaches one can today use TR-MOKE to image different spin wave modes on materials like YIG \cite{ZabelFarle13} or permalloy \cite{Wessels2016}. Improved sensitivity to spin wave dynamics in particular using optical microscopy has been demonstrated more recently by using Brillouin Light Scattering (BLS). In BLS one images not the directly reflected light from the surface of the magnetic sample but light scattered at a certain angle. The angle is determined by the interaction of the incoming photon with magnons in the sample related to the existence of magnetic excitations, like e.g. spin waves. This approach provides excellent sensitivity to magnetization dynamics, since there is very little background at the observation angle and in theory all scattered light is related to the presence of magnons. The left panel in Fig.~\ref{fig:imaging} shows an image of the envelope of a spin wave propagating from a nanoscontact located underneath the triangular shaped electrical contact. Due to the limited penetration of the light no information can be obtained underneath the electrical contact.

One way to improve the spatial resolution and to gain the ability to study buried magnetic layers is x-ray microscopy \cite{Mesler2007,Ohldag2016}, either in scanning or full field manner. These studies are usually conducted using a synchrotron x-ray source, which provides short tunable and polarized x-ray pulses between 50-100 ps duration. The absorption of circular polarized x-rays at the L-absorption resonances of the 3d transition metals like Fe, Co and Ni between 500 and 1000~eV depends strongly on the relative alignment of x-ray propagation and magnetization (X-ray Magnetic Circular Dichroism or XMCD), providing a suitable and element specific contrast mechanism for magnetic imaging. Due to the short wavelength of such x rays (1-10~nm) one is also able to obtain magnetic information on the nanoscale \cite{Stoehr2006}. If one combines x-ray microscopy with pump-probe schemes and dedicated detection electronics \cite{Acremann2006,Acremann2007,Bonetti2016}, it is then possible to follow magnetization dynamics with 10 nm spatial and 10 ps temporal resolution. One example is shown on the right hand side of Fig.~\ref{fig:imaging}, where the presence of a magnetic soliton generated by a spin-torque nano-oscillator is shown \cite{Backes2015}.	

We note that another way to image magnetization dynamics is to use scanning electron microscopy with polarization analysis (SEMPA), see for example \cite{Acremann2000,OepenHopster05}. In SEMPA a standard SEM instrument is combined with a Mott detector to identify the spin of the emitted electrons. By using an orthogonal set of Mott analyzers it is then even possible to obtain a full three dimensional map of the the magnetic behavior after excitation. However, due to the rather low efficiency of the Mott detector, the very high surface sensitivity of the detector and the vacuum requirements there are practical limitations on the types of samples that can be studied.

\section{Ultrafast Spin Transfer Torques}\label{HD}

Light is an important tool to manipulate and control magnetic properties. Advances in the development of fs optical lasers enabled radically novel ways for probing and controlling magnetism. At such sub-ps timescales the ensuing non-adiabatic dynamics is different from the LLG-based description in the rest of this chapter. It will be reviewed in detail in other chapter of this book. Briefly, the pioneering observation of sub-picosecond demagnetization in ferromagnetic nickel following fs optical laser excitation \cite{Beaurepaire96} was followed by the discovery of a wide range of laser-induced phenomena in other metallic systems, ranging from the excitation of precessional spin dynamics \cite{Ju99,Ju00,vanKampen02} to laser-induced magnetic phase transitions \cite{Thiele04,Ju04}.  More recently deterministic so-called all-optical switching by single femtosecond pulses of circularly polarized light in an increasing variety of materials and heterostructures \cite{Mangin04,Lambert14,Stanciu2007} has continued to intrigue and stimulate researchers. These effect are based on manipulating magnetic interactions such as exchange via the oscillating electric field of the exciting laser pulses. The latter mainly acts of the electron charge, effectively heating the material’s electron far above the temperature of the lattice.  This electron-light interaction, however, conserves electron spin, so it is a central question how such charge excitations can lead to a sub-ps collapse of magnetic order. 

\begin{figure}[b]
	\begin{center}	
	\includegraphics[width=75mm]{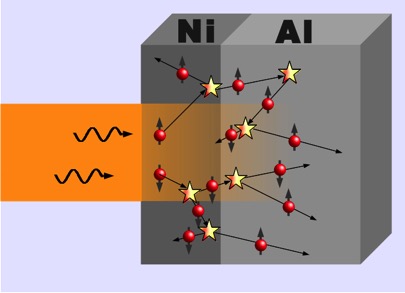}
	\caption{Schematic illustration of the fs laser excitation of superdiffusive spin currents in a Ni film. The initially ballistic hot electron motion becomes superdiffusive via scattering events altering energy and momentum of the excited electrons. The different amount of scattering for hot electrons with up and down spins leads to an effective spin current that can escape into an adjacent layer such as Al. (After Battiato {\it et al.}\cite{Battiato10}.)}
\label{FigDuerr1}
\end{center}
\end{figure}

Here we describe recent developments utilizing the spin-conserving nature of fs laser excitation of magnetic materials, i.e. laser-induced ultrafast spin currents to probe and understand their properties und attempt to utilize them for ultrafast magnetic switching at a distance. Fig.~\ref{FigDuerr1} displays the schematic excitation process of electrons from occupied electronic states below the fermi level into unoccupied states above. Theoretical modeling shows a superdiffusive spin transport of mainly majority spins away from the excitation region \cite{Battiato10}. Spin-conserving fs laser excitation leads to an excited population of electrons with both up and down spins shown in Fig.~\ref{FigDuerr1}. Spin-dependent electronic scattering processes lead to longer liftetimes for the majority spin (up) component. These spin can effectively travel a longer distance and preferrably leave the ferromagnetic layer generating a transient spin polarization in an adjacent non-magnetic layer as shown in the figure. 
 
Superdiffusive spin currents traversing a non-magnetic Au layer have been detected via non-linear second harmonic generation \cite{Melnikov11}. Ballistic Fe spins injected into a Au layer travelling close to the Au Fermi velocity arrive at the Au back interface within hundreds of femtoseconds while a diffusive component was detected at times up to 1 ps, in qualitative agreement with calculations \cite{Kaltenborn12}.  Fe layer thickness variations show that the active injection zone is an 1-2 nm thick Fe layer at the Fe/Au interface \cite{Melnikov11,Alekhin15,Melnikov15}. This implies that superdiffusive transport is of limited importance for ultrafast demagnetization of significantly thicker ferromagnetic films, an observation supported by recent demagnetization experiments in Ni films where no difference between front-side and back-side pumping was observed \cite{Schellekens13}. References\cite{Ando11,Hoffmann13} reported the detection of superdiffusive spin currents in non-magnetic layers via the inverse spin Hall effect. The use of materials displaying a large spin Hall effect allowed the detection of superdiffusive spin currents via the characteristic emitted terahertz electromagnetic pulse with a polarization determined by the transverse charge current in the spin Hall layer \cite{Kampfrath13,Seifert16}.

\begin{figure}[t]
\begin{center}
	\includegraphics[width=75mm]{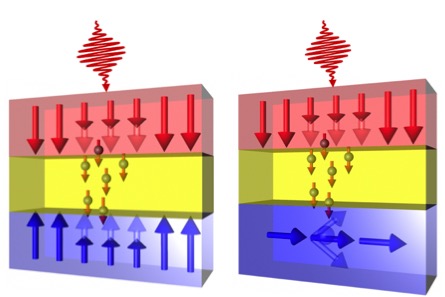}
	\caption{Schematic illustration of how superdiffusive spin currents generated by fs laser excitation of a ferromagnetic layer (red bottom layer) can be utilized to manipulate spins in a (blue) ferromagnetic layer separated by a non-magnetic spacer (yellow). (left) and (ight) panels indicate how superdiffusive spin currents can affect the magnetic moments of collinear spins and can lead to a precessional spin motion, respectively. From Hellman { {\it et al.}}~\cite{Hellman2016}.}
\label{FigDuerr2}
\end{center}
\end{figure}

In layered structures with two ferromagnetic layers with collinear magnetic moments separated by non-magnetic spacers superdiffusive spin currents preferentially excited in top magnetic layer have been found to affect the demagnetization behavior in the bottom magnetic layer \cite{Malinowski08,Rudolf12,Turgut13} (left panel of Fig.~\ref{FigDuerr2}). Replacing the spacer layer with an insulator was found to stop the superdiffusive currrents \cite{Malinowski08}. The role of transient spin accumulation at interfaces is currently not well established. The longer ballistic mean free paths for majority spin (up) electrons should lead to minority spin (down) accumulation at a ferromagnetic interface layer upon injection of an unpolarized current from an adjacent non-magnetic metal layer. However, reports for Au/Ni layers that this could even lead to ultrafast demagnetization of 15 nm thick Ni films \cite{Eschenlohr13} remain controversial \cite{Eschenlohr14,Khorsand14}. However a majority spin transmission through Pt(30nm)/[Co/Pt] (6.5nm)/Cu (100-200nm) heterostructures was observed via transient spin accumulation in the Cu layer following fs laser heating of the Pt layer \cite{Choi14}. We finally point out that the insertion of a tunneling barrier between two ferromagnetic layers enables a new control mechanism via magnetic tunnel junctions. He {\it et al.} \cite{He13} observed that spin tunneling through MgO spacers could influence the ultrafast demagnetization of adjacent CoFeB layers. A first attempt at contoling the femtosecond demagnetization in CoFeB-based magnetic tunnel junctions via tuning the voltage applied to the junction was reported in \cite{Savoini14}.

One of the most typical spintronics devices is the spin-transfer torque magnetic random access memory, where a spin current is used to exert a torque on a magnetic bit ultimately switching its direction. Under typical operation conditions, near equilibrium, large enough spin currents can only be generated in nanopillar devices. The use of strong, ultrashort non-equilibrium spin currents could open up new ways for spin transfer torque switching. Schellekens et al. \cite{Schellekens14} and Choi et al. \cite{Choi14} simultaneously demonstrated spin torque induced precession dynamics driven by superdiffusive spin currents. These experiments utilize two ferromagnetic layers with in-plane and out-of-plane magnetization separated by Cu and Pt \cite{Schellekens14} and Cu \cite{Choi14} spacer layers of varying thickness. Although the induced precession angles are still small due to the limited amount of observed spin angular momentum transfer of several percent between the orthogonally magnetized layers, such experiments represent a unique tool to actually quantify and consequently optimize angular momentum transfer through interfaces.

A key ingredient in all-optical switching presented in more detail in the chapter by T. Rasing {\em et. al} in this book, is the dramatic reduction of the sample magnetization via ultrafast demagnetization. This allows angular momentum exchange between magnetic subsystems to take over as clearly demonstrated by ultrafast element-specific x-ray magnetometry in GdFeCo films \cite{Radu11}. The emergence of a ferromagnetic transient state out of the antiferromagnetically aligned transition metal and rare earth magnetic subsystems bears this fingerprint of non-local angular momentum exchange \cite{Radu11}. The question arises whether demagnetization accompanied by superdiffusive spin currents can also enable magnetic switching. Graves {\it et al.} \cite{Graves13} found such a mechanism for GdFeCo using ultrafast x-ray scattering at novel x-ray free electron laser facilities. The basis of this study was the chemical segregation observed in GdFeCo films into Gd-rich and Fe-rich nanoregions on a ~10 nm lengthscale. Following ultrafast demagnetization of both the Gd 4f and Fe 3d magnetic sublattice a reversal of the Gd 4f moments in Gd-rich nanoregions was observed. This led to a local ferromagnetic state with parallel alignment of Fe 3d and Gd 4f moments. The authors also measured the net spin angular momentum transfer across ~10 nm distances and showed that it is compatible with Fe 3d spins from Fe-rich regions being transported laterally into Gd-rich areas to accumulate there until the originally antiparallel Gd 4d magnetization is reversed \cite{Graves13}. These measurements lead the way to study the effects of ultrafast spin transport over nanoscale dimensions. Although currently only the action on local magnetic moments can be probed \cite{Graves13} future advances in x-ray nano-spectroscopy offer the unique opportunity to also determine the effects of non-local transport currents on local valence level populations \cite{Kukreja2015}.

\section{Summary and outlook}
Modern magnetism and magneto-dynamics have seen significant progress in the last two decades. The theoretical discovery and experimental observation of spin-torque opened up new frontiers for fully localized control of nanomagnetic objects. This provided a useful means of writing a magnetic bit in high-density arrangement such as silicon-integrated STT-MRAM. Spin-torque also brought about a new type of dynamics involving negative damping, opening up the world of spin-current driven nonlinear magnetic oscillators, new ways of propagating and manipulating spin-waves, and new nonlinear magnetic condensates. Much of the nonlinear properties of such magnetodynamics is still being explored at the time of this writing. For such exploration the new metrologies developed during the last two decades have become very important, those of XMCD-based spin-resolved and time-resloved microscopy with sub-20nm resolution has enabled direct observation of many such dynamics states, accelerating in many cases our understanding of such complex systems. Beyond transport current-induced spin dynamics, light-interaction with magnetic materials have also been shown to induce significant non-equilibrium spin-angular momentum change of the electronic systems in such materials. These have been demonstrated to result in ultra-fast magnetic switching, with the electronic process completing as fast as a few hundred femto-seconds. A deeper understanding of the materials and electronic physics is well under way, and will hopefully be addressing future technology needs in the ultra-fast realm. Like many other branches of scientific inquiry, the field of magneto-dynamics has been rejuvenated by an influx of new ideas and new scientific methods as well as the development of ever-so-more advanced measurement capability. Whereas in the recent past such research and development effort was often centered around magnetic storage and recording, the frontier is now broader. Magnetism being fundamentally nonlinear and rich in its dynamic behaviors continues to fascinate us all, both in revealing new scientific insights, and in providing new technologies for the future.

\section{Acknowledgment}
A.D.K. acknowledges support by the National Science Foundation under award DMR-1610416.

\bibliographystyle{spphys}

\end{document}